\documentclass{ws-ijbc}
\usepackage{ws-rotating}     
\usepackage{graphicx}

\begin{document}

\catchline{}{}{}{}{} 
\markboth{Author's Name}{Paper Title}

\title{Chimera states in networks of nonlocally coupled Hindmarsh-Rose neuron models}

\author{Johanne Hizanidis}
\address{National Center for Scientific Research ``Demokritos''\\
Athens, Greece\\
ioanna@chem.demokritos.gr
}

\author{Vasilis Kanas}
\address{Department of Electrical and Computer Engineering\\
University of Patras, Patras, Greece\\
vaskanas@upatras.gr
}

\author{Anastasios Bezerianos}
\address{Cognitive Engineering Lab, Singapore Institute for Neuroengineering (SINAPSE)\\
National University of Singapore, Singapore\\
bezer@upatras.gr
}

\author{Tassos Bountis}
\address{Department of Mathematics and\\ Center of Research and Applications of Nonlinear Systems\\
University of Patras, Patras, Greece\\
tassos50@otenet.gr
}

\maketitle

\begin{history}
\received{(to be inserted by publisher)}
\end{history}

\begin{abstract}

We have identified the occurrence of chimera states for various coupling schemes in networks of two-dimensional and three-dimensional Hindmarsh-Rose oscillators, which represent realistic models of neuronal ensembles. This result, together with recent studies on multiple chimera states in nonlocally coupled FitzHugh-Nagumo oscillators, provide strong evidence that the phenomenon of chimeras may indeed be relevant in neuroscience applications. Moreover, our work verifies the existence of chimera states in coupled bistable elements, whereas to date chimeras were known to arise in models possessing a single stable limit cycle. Finally, we have identified an interesting class of mixed oscillatory states, in which desynchronized neurons are uniformly interspersed among the remaining ones that are either stationary or oscillate in synchronized motion.
 
\end{abstract}


\section{Introduction}

About ten years ago, a peculiar dynamical phenomenon was discovered in populations of identical phase oscillators:
Under nonlocal symmetric coupling, the coexistence of coherent (synchronized) and incoherent oscillators was observed \cite{KUR02}. 
This highly counterintuitive phenomenon was given the name chimera state after the Greek mythological creature made up of different animals \cite{ABR04}.
Since then the study of chimera states has been the focus of extensive research in a wide number of models, from Kuramoto phase oscillators \cite{ABR08,LAI12} to periodic and chaotic maps \cite{OME11}, as well as Stuart-Landau oscillators \cite{LAI10}.
The first experimental evidence of chimera states was found in populations of coupled chemical oscillators as well as in optical coupled-map lattices realized by liquid-crystal light modulators \cite{TIN12,HAG12}.
Recently, moreover, Martens and coauthors showed that chimeras emerge naturally from a competition between two antagonistic synchronization patterns in a mechanical experiment involving two subpopulations of identical metronomes
coupled in a hierarchical network \cite{MAR13}.

In the context of neuroscience, a similar effort has been undertaken by several groups, since it is believed that chimera states might explain the phenomenon of unihemispheric sleep observed
in many birds and dolphins which sleep with one eye open, meaning that one hemisphere of the brain is synchronouns whereas the other is asynchronous \cite{RAT00}. The purpose of this paper is to make a contribution in this direction, by identifying for the first time a variety of single and multi-chimera states in networks of non-locally coupled neurons represented by Hindmarsh--Rose oscillators.   

Recently, multi-chimera states were discovered on a ring of nonlocally coupled
FitzHugh-Nagumo (FHN) oscillators \cite{OME13}. The FHN model is a  2--dimensional (2D) simplification of the physiologically realistic Hodgkin-Huxley model \cite{HOD52} and is therefore computationally a lot
simpler to handle. However, it fails to reproduce several important dynamical behaviors shown by real neurons, like rapid firing or regular and chaotic bursting. This can be overcome by replacing the FHN with another well-known more realistic model
for single neuron dynamics, the Hindmarsh-Rose (HR) model \cite{HIN82,HIN84}, which we will be used throughout this work both in its 2D and 3D versions.

In Section 2 of this paper, we first treat the case of 2D-HR oscillators represented by two first order ordinary differential equations (ODEs) describing the interaction of a membrane potential and a single variable related to ionic currents across the membrane under periodic boundary conditions. We review the dynamics in the 2D plane in terms of its fixed points and limit cycles, coupling each of the $N$ oscillators to $2R<N$ nearest neighbors symmetrically on both sides, in the manner adopted in \cite{OME13}, through which chimeras were discovered in FHN oscillators. 

We identify parameter values for which chimeras appear in this setting and note the variety of oscillating patterns that are possible due to the bistability features of the 2D model. In particular, we identify a new ``mixed oscillatory state'' (MOS), in which the desynchronized neurons are uniformly distributed among those attracted by a stable stationary state. Furthermore, we also discover chimera--like patterns in the more natural setting where only the membrane potential variables are coupled with $2R$ of the same type.

Next, we turn in Section 3 to the more realistic 3D-HR model where a third variable is added representing a slowly varying current, through which the system can also exhibit bursting modes. Here, we choose a different type of coupling where the two first variables are coupled symmetrically to $2R$ of their own kind and observe only states of complete synchronization as well as MOS in which desynchronized oscillators are interspersed among neurons that oscillate in synchrony. However, when coupling is allowed only among the first (membrane potential) variables chimera states are discovered in cases where spiking occurs within sufficiently long bursting intervals. Finally, the paper ends with our conclusions in Section 4.

\section{Two--dimensional HR models}
Following the particular type of setting proposed in \cite{OME13} we consider $N$ nonlocally coupled Hindmarsh-Rose oscillators, where the interconnections between neurons exist with $R$ nearest neighbors only on either side as follows:
\begin{eqnarray}
\dot x_k &=& y_k-x_k^3+3x^2+J+\frac{\sigma_x}{2R}\sum_{j=k-R}^{j=k+R} [b_{xx}(x_j-x_k)+b_{xy}(y_j-y_k)] \label{eq:01}
\\
\dot y_k &=& 1-5x_k^2-y_k+
\frac{\sigma_y}{2R}\sum_{j=k-R}^{j=k+R} [b_{yx}(x_j-x_k)+b_{yy}(y_j-y_k)]. \label{eq:02}  
 \end{eqnarray}
In the above equations $x_k$ is the membrane potential of the $k$-th neuron, $y_k$ represents various physical
quantities related to electrical conductances of the relevant ion currents across the membrane, $a=c=1$, $b=3$ and $d=5$ are constant parameters, and $J=0$ is the external stimulus current.  Each oscillator is coupled with its $R>0$ nearest neighbors on both sides with coupling strengths $\sigma_x, \sigma_y>0$. This induces nonlocality in the 
form of a ring topology established by considering periodic boundary conditions for our systems of ODEs. 
As in \cite{OME13}, our system contains not only direct $x-x$ and $y-y$ coupling, but also cross-coupling
between variables $x$ and $y$. This feature is modeled by a rotational coupling matrix:
\[ B= \left(\begin{array}{cc}
b_{xx} & b_{xy}  \\
b_{yx} & b_{yy}   
\end{array}\right) = \left(\begin{array}{cc}
cos\phi & sin\phi  \\
-sin\phi & cos\phi   
\end{array}\right)\] 
depending on a coupling phase $\phi$.
In what follows, we study the collective behavior of the above system and investigate, in particular,  the existence of chimera states in relation to the values of all network parameters: $N$, $R$, $\phi$, $\sigma_x$ and $\sigma_y$.
More specifically, we consider two cases: ($\mathrm{I}$) direct and cross-coupling of equal strength in both variables ($\sigma_x=\sigma_y$) and ($\mathrm{II}$) direct coupling in the $x$ variable only ($\sigma_y=0$, $ \phi=0$). Similarly to \cite{OME13} we shall use initial conditions randomly distributed on the unit circle ($x^2+y^2=1$).

\begin{figure}[h]
\begin{center}
\vspace{1.0cm}
\includegraphics[width=0.7\textwidth]{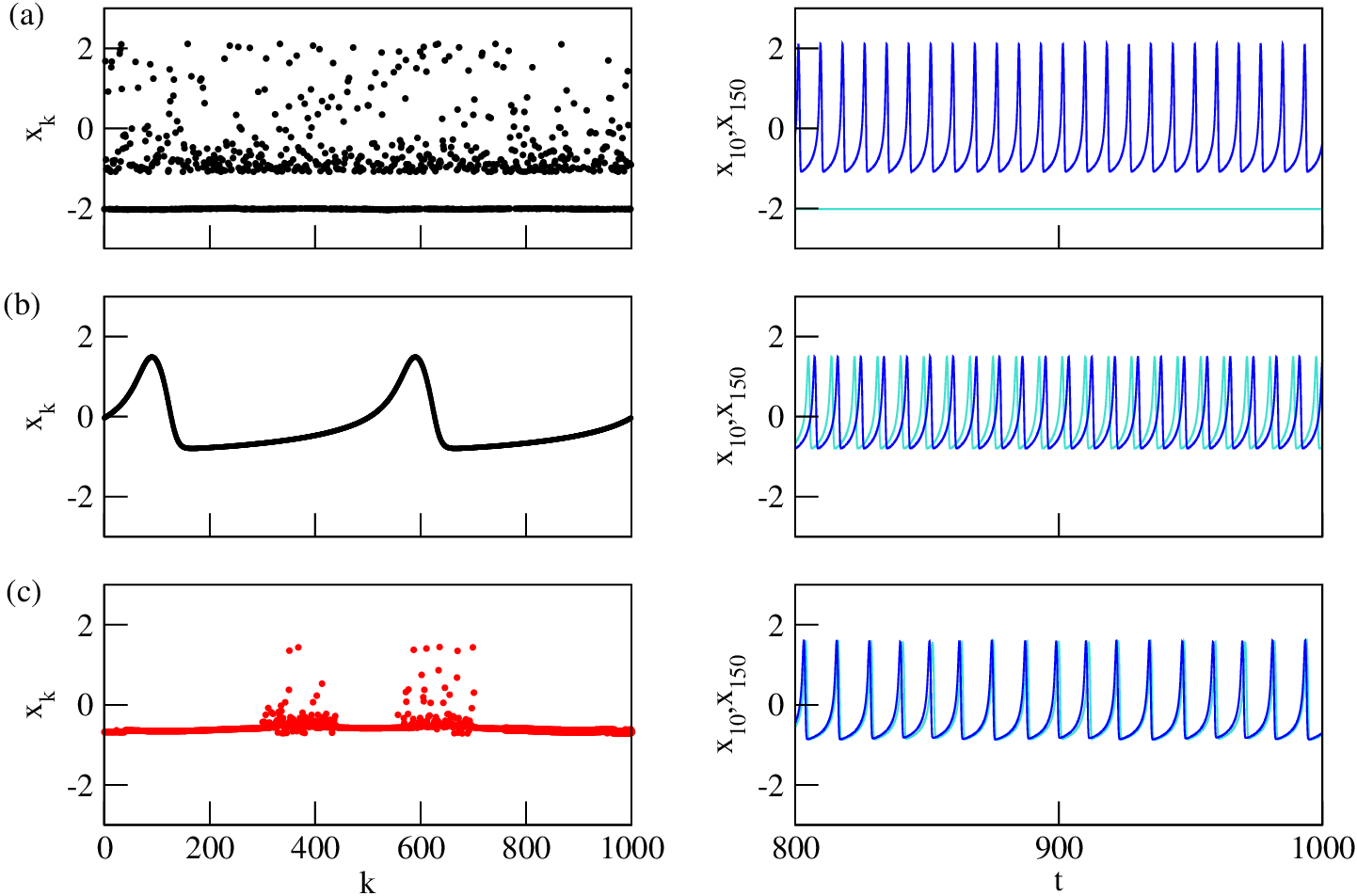}
\end{center}
\caption{Snapshots of the variable $x_k$ at $t=3000$ of Eqs.~(\ref{eq:01}),(\ref{eq:02}) (left)
and selected time series (right)
for $\sigma_x=\sigma_y=0.1$. (a) $\phi=-\pi$, (b) $\phi=-\pi/4$ and (c) $\phi=0$. $N=1000$
and $R=350$.}
\label{fig:fig1}
\end{figure}

Typical spatial patterns for case ($\mathrm{I}$) are shown on the left panel of Fig.~\ref{fig:fig1}, where the $x$ variable is plotted over the index number $k$ at a time snapshot chosen after a sufficiently long simulation of the system. In this figure the effect of different values of the phase $\phi$ is demonstrated while the number of oscillators $N$ and their nearest neighbors $R$, as well as the coupling strengths $\sigma_x=\sigma_y$ are kept constant. For example, for $\phi=-\pi$ (Fig.~\ref{fig:fig1}(a)) an interesting novel type of dynamics is observed that we shall call ``mixed oscillatory state'' (MOS), whereby nearly half of the $x_k$ are attracted to a fixed point (at this snapshot they are all at a value near $-2$), while the other half are oscillating interspersed among the stationary ones. From the respective time series (Fig.~\ref{fig:fig1}(a), right) it is clear that the former correspond to spiking neurons whereas the latter to quiescent ones.

\begin{figure}[h]
\begin{center}
\includegraphics[width=0.4\textwidth]{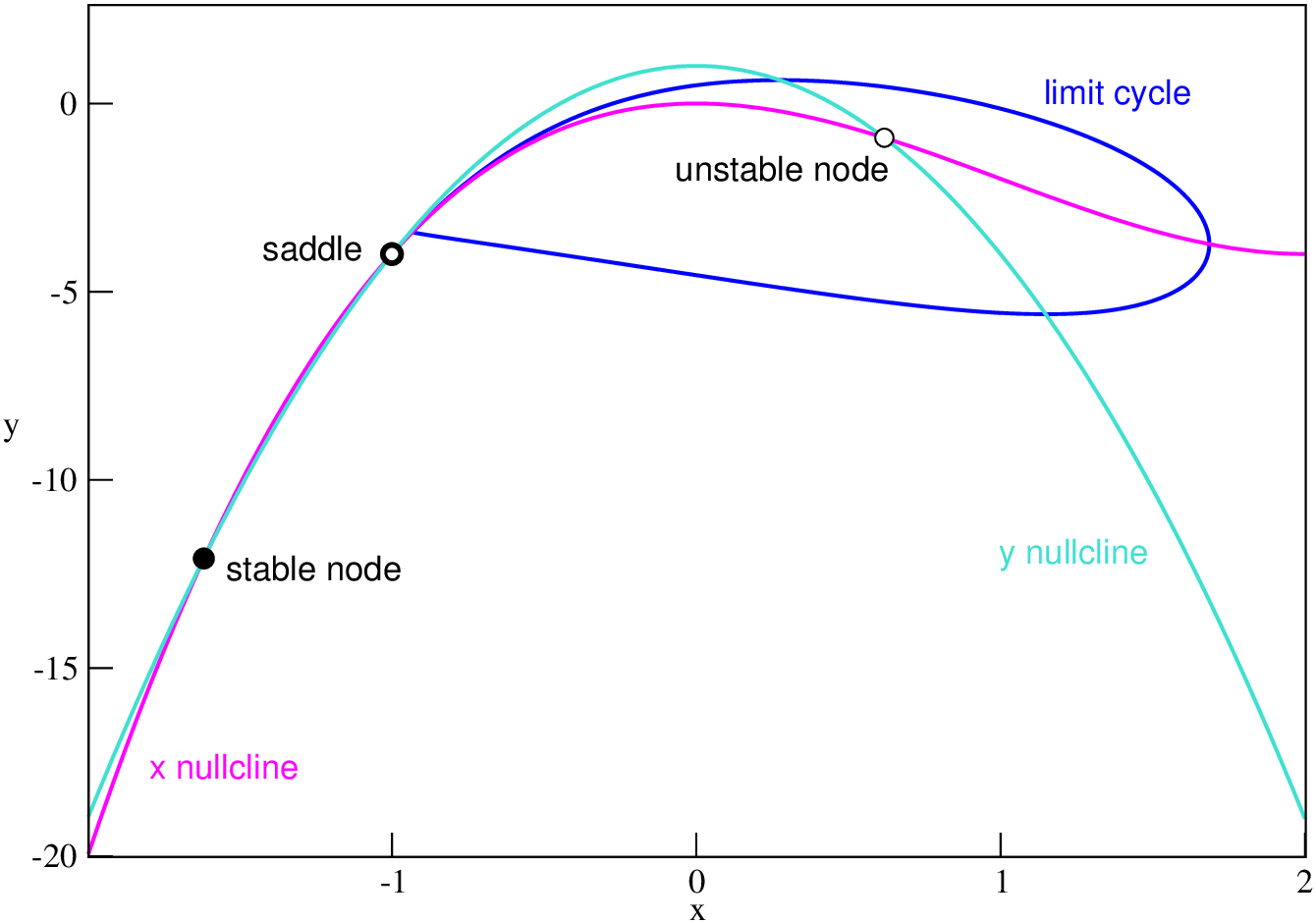}
\end{center}
\caption{Nullclines of Eqs. (1) and (2) for $J=0$. Three fixed points coexist with a stable limit cycle.}
\label{fig:fig2}
\end{figure}

This interesting MOS phenomenon is due to the fact that, in the standard parameter values we have chosen, the uncoupled HR oscillators are characterized by the property of {\em bistability}. Clearly, as shown in the phase portrait of Fig.~\ref{fig:fig2}, each oscillator possesses three fixed points: The leftmost fixed point is a stable node corresponding to the resting state of the neuron while the other two correspond to a saddle point and an unstable node and are therefore repelling. 
For $J=0$ (which is the case here) a stable limit cycle also exists which attracts many of the neurons into oscillatory motion, rendering the system bistable and producing the dynamics observed in Fig.~\ref{fig:fig1}(a). Now, when a positive current $J>0$ is applied, the $x$-nullcline is lowered such that the saddle point and the stable node collide and finally vanish. In this case the full system enters a stable limit cycle associated with typical spiking behaviour. Similar complex patterns including MOS and chimeras have been observed in this regime as well.

For $\phi=-\pi/4$, on the other hand, there is a ``shift'' in the dynamics of the individual neurons into the spiking regime, as seen in Fig.~\ref{fig:fig1}(b) (right). The corresponding spatial pattern has a wave-like form of period 2.
Then, for $\phi=0$ a classical chimera state with two incoherent domains is observed ((Fig.~\ref{fig:fig1}(c), left).
Diagonal coupling ($b_{yx}=b_{xy}=0$, $b_{xx}=b_{yy}=1$) is, therefore, identified as the necessary condition to achieve chimera states. By contrast, it is interesting to note that in nonlocally coupled Fitz-Hugh Nagumo oscillators \cite{OME13} it has been shown both analytically and numerically that chimera states occur for {\it off-diagonal} coupling. 
 
\begin{figure}[h]
\begin{center}
\includegraphics[width=0.5\textwidth]{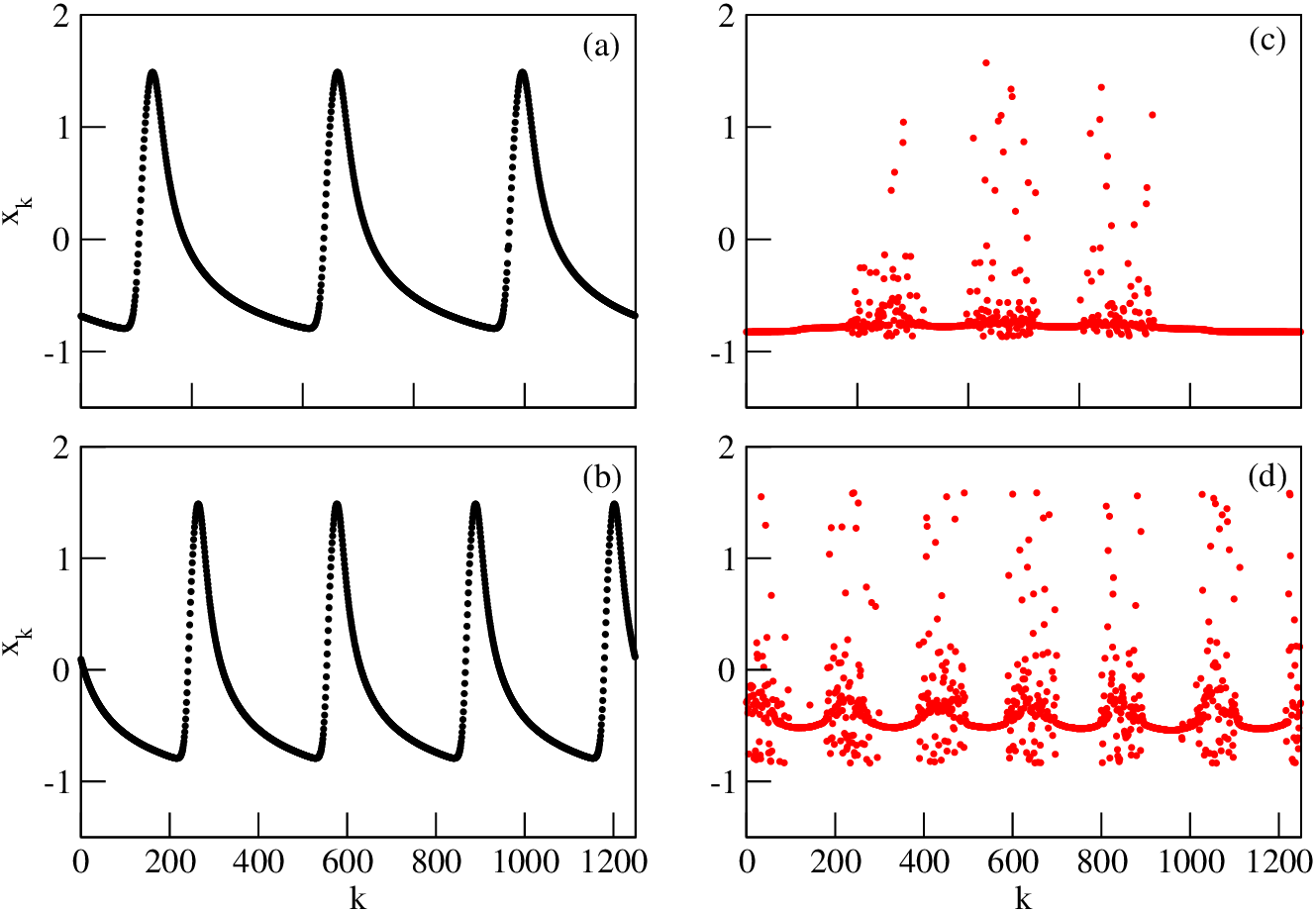}
\end{center}
\caption{Snapshots of the variable $x_k$ of Eqs.~(\ref{eq:01}),(\ref{eq:02}) at $t=3000$ for 
$\sigma_x=\sigma_y$. $N=1000, R=250$ (top) and $N=1250, R=250$ (bottom) whereas $\phi=-\pi/4$ (left) and $\phi=0$ (right).
In (a) and (b) $\sigma_x=0.1$, in (c) $\sigma_x=0.03$, and in (d) $\sigma_x=0.049$.}
\label{fig:fig3}
\end{figure}

By decreasing the range of coupling $R$ and increasing the system size $N$,  chimera states occur with multiple domains of incoherence and coherence for $\phi=0$ (Fig.~\ref{fig:fig3}(c,d)), and, accordingly, periodic spatial patterns with larger wave numbers arise for $\phi=-\pi/4$, as seen in Fig.~\ref{fig:fig3}(a,b).
This is in agreement with previous works reported in \cite{OME11,OME13}.

\begin{figure}[h]
\begin{center}
\includegraphics[width=0.4\textwidth]{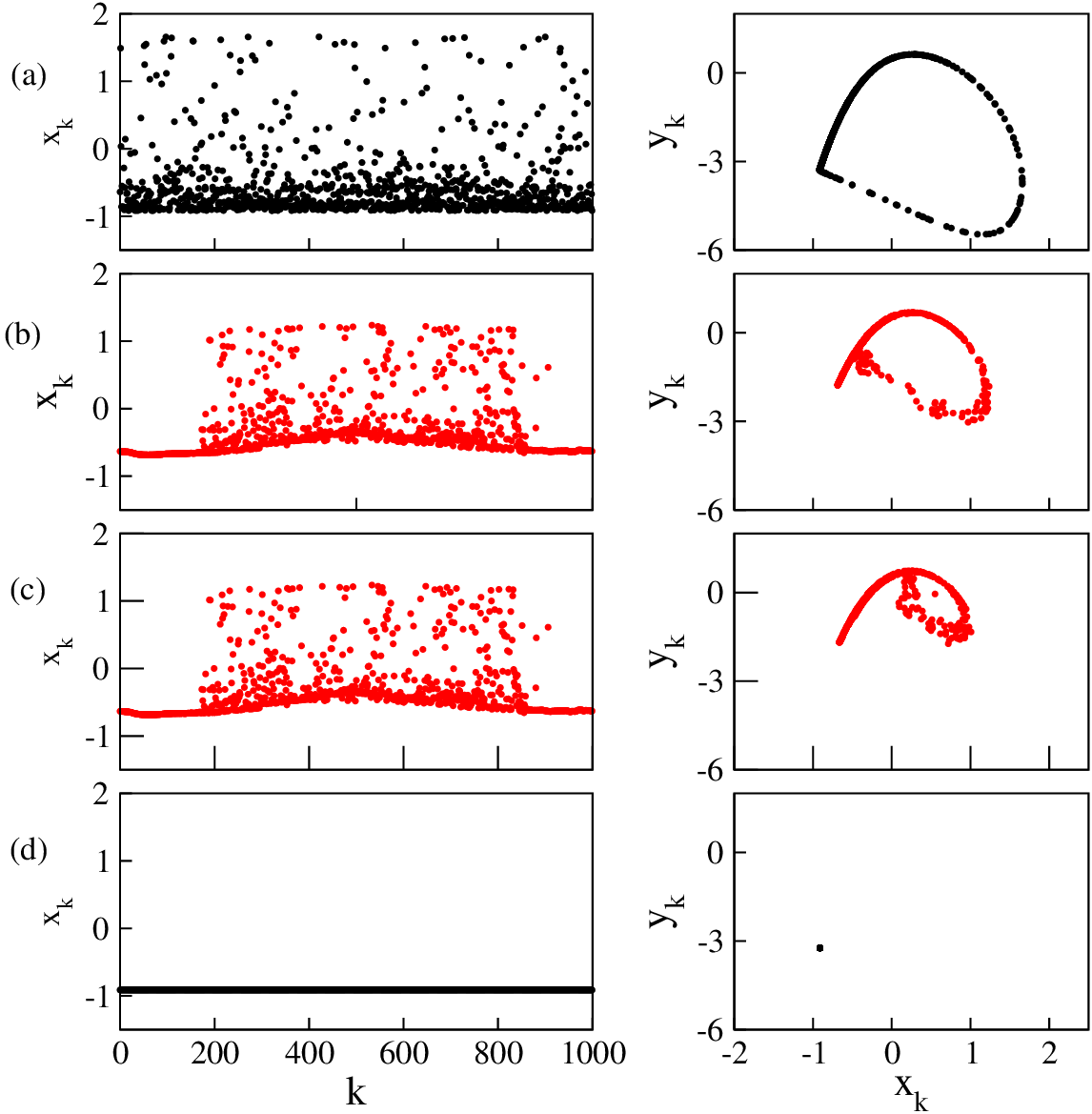}
\end{center}
\caption{Snapshots of the variable $x_k$  
(left) and in the $(x_k, y_k)$-plane (right) of Eqs.~(\ref{eq:01}),(\ref{eq:02}) at $t=3000$
for $\sigma_y=0$, $N=1000$ and $R=350$. (a) $\sigma_x=0.01$, (b) $\sigma_x=0.4$, 
(c) $\sigma_x=0.6$ and (d) $\sigma_x=1.05$.}
\label{fig:fig4}
\end{figure}

Next we consider the case ($\mathrm{II}$) where the coupling term is restricted to the $x$ variable. This case is important since incorporating the coupling
in the voltage membrane ($x$) alone is more realistic from a  biophysiological point of view.
In Fig.~\ref{fig:fig4} spatial plots (left) and the corresponding $(x_k,y_k)$-plane (right) 
for increasing coupling strength are shown.
Chimera states (Fig.~\ref{fig:fig4} (b,c)) are observed as an intermediate pattern between
desynchronization (Fig.~\ref{fig:fig4} (a)) and complete synchronization (Fig.~\ref{fig:fig4} (d)).

\section{Three--dimensional HR models}

In order to complete our study of the Hindmarsh-Rose model we shall consider, in this section, its three-dimensional version. The corresponding equations read:

\begin{eqnarray}
 \dot x_k &=& y_k-a{x_k}^3+b{x_k}^2+J-z+\frac{\sigma_x}{2R}\sum_{j=k-R}^{j=k+R} \left(x_j-x_k\right)
 \label{eq:03}\\
 \dot y_k &=& c-d{x_k}^2-y_k +\frac{\sigma_y}{2R}\sum_{j=k-R}^{j=k+R} \left(y_j-y_k\right) \label{eq:04}
 \\
 \dot z_k &=& r (s(x_k-x_0)-z_k) \label{eq:05} 
 \end{eqnarray}

The extra variable $z$ represents a slowly varying current, which changes the applied current $J$ to $J-z$ and guarantees firing frequency adaptation (governed by the parameter $s$) as well as the ability to produce  typical bursting modes, which the 2D model cannot reproduce. Parameter $b$ controls the transition between spiking and bursting, parameter $r$ determines the spiking frequency (in the spiking regime) and the number of spikes per bursting (in the bursting regime), while $x_0$ sets the resting potential of the system. The parameters of the fast $x-y$ system are set to the same values used in the two-dimensional version ($a=c=1$, $d=5$) and typical values are used for the parameters of the $z$-equation 
($r=0.01$, $s=4$, $x_0=-1.6$).

The 3D Hindmarsh-Rose model exhibits a rich variety of bifurcation scenarios in the $b-J$ parameter plane \cite{STO08}. Thus, we prepare all nodes in the spiking regime (with corresponding parameter values $b=3$ and $J=5$)
and, as in Section 2, we use initial conditions randomly distributed on the unit sphere ($x^2+y^2+z^2=1$). 

\begin{figure}[]
\begin{center}
\includegraphics[width=0.5\textwidth]{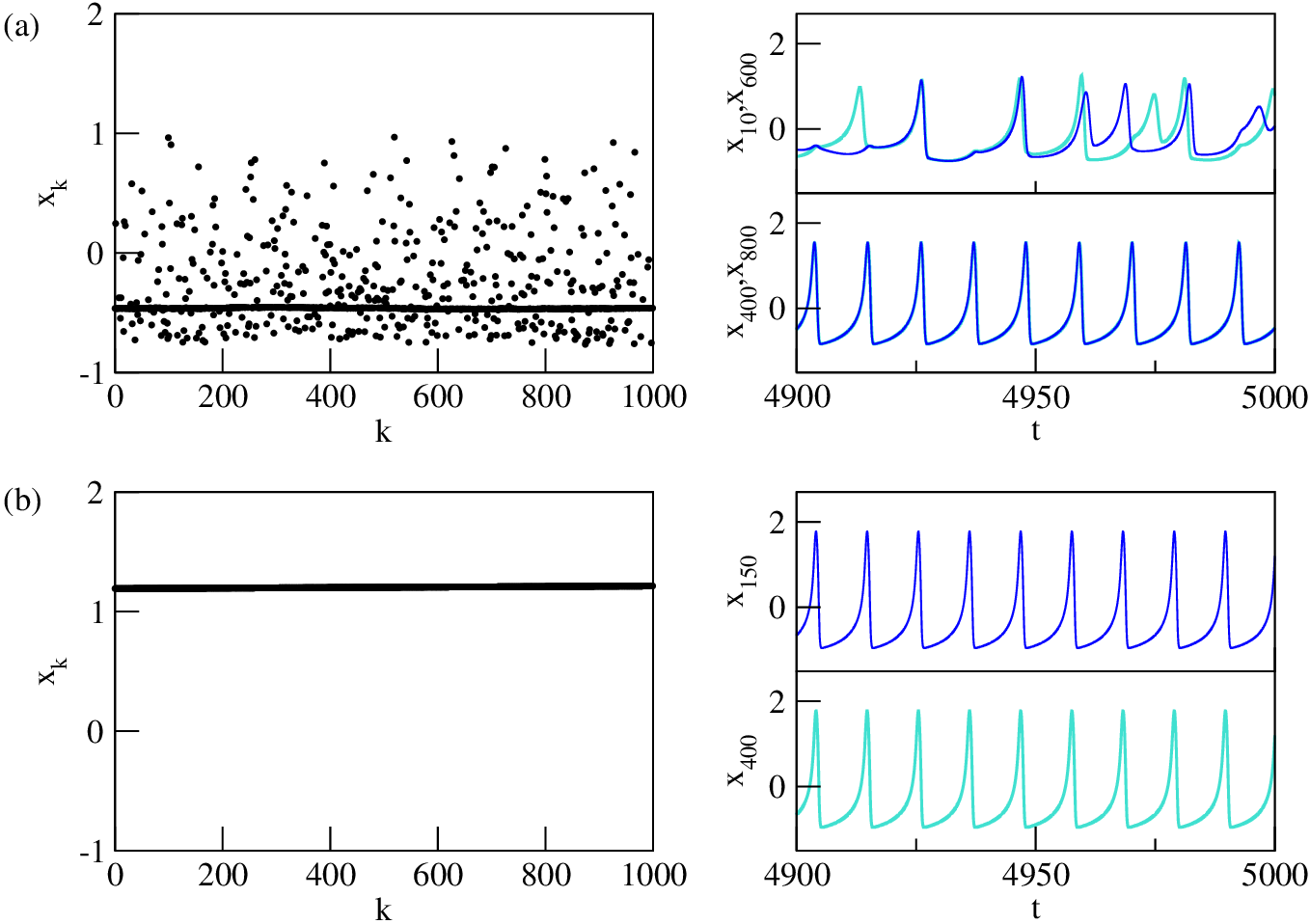}
\end{center}
\caption{Snapshots of the variable $x_k$ of Eqs.~(\ref{eq:03})-(\ref{eq:05}) at $t=5000$ (left) and selected time series (right) for: (a) $\sigma_x=\sigma_y=0.14$ and (b)  $\sigma_x=\sigma_y=0.29$. $N=1000$ and $R=350$.}
\label{fig:fig5}
\end{figure}

First let us consider direct coupling in both variables $x$ and $y$ and vary the value of the equal coupling strengths $\sigma_x=\sigma_y$, while $N$ and $R$ are kept constant.  Naturally, the interaction of $N$ spiking neurons will lead to a change in their dynamics, as we discuss in what follows. For low values of $\sigma$ we observe the occurrence of a type of MOS where nearly half of the neurons spike regularly in a synchronous fashion, while the rest are unsynchronized and spike in an irregular fashion.
This is illustrated in the respective time series in the right panel of Fig.~\ref{fig:fig5}(a)
At higher values of the coupling strength the the system is fully synchronized (Fig.~\ref{fig:fig5}(b)).

\begin{figure}[h]
\begin{center}
\includegraphics[width=0.5\textwidth]{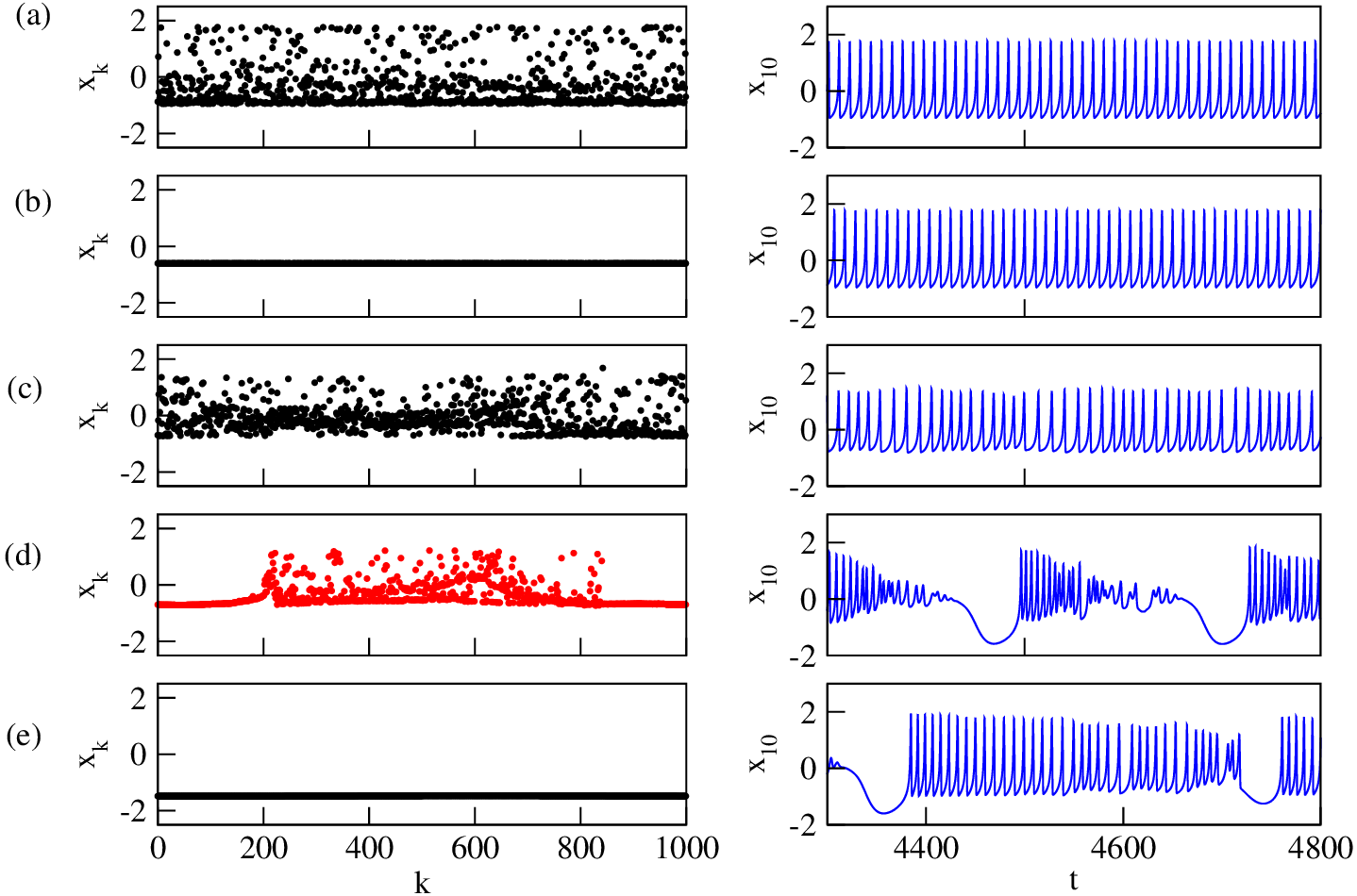}
\end{center}
\caption{Snapshots of the variable $x_k$ of Eqs.~(\ref{eq:03})-(\ref{eq:05}) at $t=5000$ (left) and selected time series
for $\sigma_y=0$, $N=1000$ and $R=350$.
(a) $\sigma_x=0.005$, (b) $\sigma_x=0.02$, $\sigma_x=0.17$, (d) $\sigma_x=0.47$, and (e) $\sigma_x=1.0$.
}
\label{fig:fig6}
\end{figure}

Next we check the case of coupling in the $x$ variable alone ($\sigma_x=\sigma$, $\sigma_y=0$). Figure~\ref{fig:fig6} displays characteristic synchronization patterns obtained when we increase $\sigma_x$ (left panel) and selected time series (right panel).
At low values of the coupling strength all neurons remain in the regular spiking 
regime and desynchronization alternates with complete synchronization as $\sigma_x$ increases (Fig.~\ref{fig:fig6}(a,b,c)).
For intermediate values of the coupling strength, chimera states with one incoherent domain are 
to be observed. These are associated with a change in the dynamics of the individual neurons, which
now produce irregular bursts (Fig.~\ref{fig:fig6}(d)). The number of spikes in each burst increases at higher $\sigma_x$ values and the system is again fully synchronized (Fig.~\ref{fig:fig6}(e)). 
Extensive simulations show that the chimera states disappear and reappear again by varying $\sigma_x$ 
which is most likely due to the system's multistability and sensitive
dependence on initial conditions. 

\section{Conclusions}

In this paper, we have identified the occurrence of chimera states for various coupling schemes 
in networks of 2D and 3D Hindmarsh-Rose models. This, together with recent reports on multiple chimera states in
nonlocally coupled FitzHugh-Nagumo oscillators, provide strong evidence that this counterintuitive phenomenon is very relevant as far as neuroscience applications are concerned. 

Chimera states are strongly related to the phenomenon of synchronization.
During the last years, synchronization phenomena have been intensely studied
in the framework of complex systems \cite{ARE08}. Moreover, it is  
a well-established fact the key ingredient for the occurrence
of chimera states is nonlocal coupling. The human brain is an excellent example 
of a complex system where nonlocal connectivity
is compatible with reality. Therefore, the study of chimera states in systems modelling neuron dynamics is both significant and relevant as far as applications are concerned. 

Moreover, the present work is also important from a theoretical point of view, since it verifies the existence of chimera states in coupled bistable elements, while, up to now, it was known to arise in oscillator models possessing a single attracting state of the limit cycle type. Finally, we have identified a novel type of mixed oscillatory state (MOS), in which desynchronized neurons are interspersed among those that are either stationary or oscillate in synchrony. As a continuation of this work, it is very interesting to see whether chimeras and MOS states appear also in networks of {\it populations} of Hindmarsh-Rose oscillators, which are currently under investigation.

\section{Acknowledgments}

The authors acknowledge support by the European
Union (European Social Fund – ESF) and Greek national
funds through the Operational Program ``Education and
Lifelong Learning'' of the National Strategic Reference Framework (NSRF) -
Research Funding Program: THALES. Investing in knowledge society through the
European Social Fund. Funding was also provided by NINDS R01-40596.

\end{document}